\documentclass[AMA,STIX1COL]{WileyNJD-v2}

\articletype{Research Article}%

\received{26 April 2016}
\revised{6 June 2016}
\accepted{6 June 2016}

\usepackage{svg}
\usepackage{tikz}
\usepackage{graphicx}
\usepackage{caption}
\usepackage{subcaption}
\usepackage{svg}
\usepackage{url}

\usepackage{xcolor}

\usepackage{color}
\usepackage{tabularray}
\definecolor{JapaneseLaurel}{rgb}{0,0.501,0}
\definecolor{WebOrange}{rgb}{1,0.647,0}

\definecolor{codegreen}{rgb}{0,0.6,0}
\definecolor{codegray}{rgb}{0.5,0.5,0.5}
\definecolor{codepurple}{rgb}{0.58,0,0.82}
\definecolor{backcolour}{rgb}{0.95,0.95,0.92}

\newcommand{\ballnumber}[1]{\tikz[baseline=(myanchor.base)] \node[circle,fill=.,inner sep=1pt] (myanchor) {\color{-.}\bfseries\footnotesize #1};}

\usepackage{parcolumns}
\usepackage{float}
\usepackage{listings}
\usepackage{xcolor}
\usepackage{amssymb}%
\usepackage{pifont}%
\newcommand{\cmark}{\ding{51}}%
\newcommand{\xmark}{\ding{55}}%

\begin{document}

\title{CloudSim Express: A Novel Framework for Rapid Low Code Simulation of Cloud Computing Environments}

\author[1]{Tharindu B. Hewage}

\author[2]{Shashikant Ilager}

\author[1]{Maria A. Rodriguez}

\author[1]{Rajkumar Buyya}

\authormark{Hewage, Ilager, Rodriguez, and Buyya}

\address[1]{\orgdiv{The Cloud Computing and Distributed Systems (CLOUDS) Laboratory}, \orgname{School of Computing and Information Systems}, \orgaddress{\state{The University of Melbourne}, \country{Australia}}}

\address[2]{
\orgaddress{\state{Vienna University of Technology (TU Wien)}, \country{Vienna, Austria}}}

\corres{Tharindu B. Hewage. \email{tsaryakarahe@student.unimelb.edu.au}}

\abstract[Abstract]{Cloud computing environment simulators enable cost-effective experimentation of novel infrastructure designs and management approaches by avoiding significant costs incurred from repetitive deployments in real Cloud platforms. However, widely used Cloud environment simulators compromise on usability due to complexities in design and configuration, along with the added overhead of programming language expertise. Existing approaches attempting to reduce this overhead, such as script-based simulators and Graphical User Interface (GUI) based simulators, often compromise on the extensibility of the simulator. Simulator extensibility allows for customization at a fine-grained level, thus reducing it significantly affects flexibility in creating simulations.
To address these challenges, we propose an architectural framework to enable human-readable script-based simulations in existing Cloud environment simulators while minimizing the impact on simulator extensibility. We implement the proposed framework for the widely used Cloud environment simulator, the CloudSim toolkit, and compare it against state-of-the-art baselines using a practical use case. The resulting framework, called \textit{CloudSim Express}, achieves extensible simulations while surpassing baselines with over a $ 71.43$\% reduction in code complexity and an $89.42$\% reduction in lines of code.}

\keywords{Cloud Computing, Modelling and Simulation, CloudSim, Programming Productivity}

\maketitle

\renewcommand\thefootnote{}

\renewcommand\thefootnote{\fnsymbol{footnote}}
\setcounter{footnote}{1}

\section{Introduction}
\label{section-introduction}

\par Cloud computing evolves at a fast rate, facilitating a wide variety of applications and services \cite{david2022futureofcloud2027}. This leads to the continuing development of novel infrastructure designs and management approaches. However, validating such approaches in actual Cloud environments incurs significant costs, time, and effort due to repetitive testing. Consequently, Cloud environment simulators are used to explore, analyze, and stress such approaches before deploying them in actual environments. Simulations allow for the validation of strategies through repetitive experiments with fewer resources \cite{mansouri2020cloussimulatorsurvey}. For example, novel resource allocation strategies in hyper-scale Cloud environments need to be iteratively benchmarked. Performing these tests in a real-world setting would result in significant financial costs, development time, and effort compared to less resource-intensive simulators \cite{mansouri2020cloussimulatorsurvey}. Established Cloud environment simulators are generally preferred for simulated scenarios, primarily due to their feature availability \cite{li2012dartcsim}. However, existing simulators are bound to a specific programming language and its ecosystem. This require designing simulation scenarios via code, performing various configurations, and re-compiling the code upon changing the simulation \cite{calheiros2011cloudsim, malik2014cloudnetsim++, damian2018score}. Furthermore, a simulation involves maintaining a significant amount of code, making it complex to find defects unless appropriate programming practices are followed \cite{sadowski2018moderncodereviewgoogle}. 

\par On the other hand, script-based approaches are preferred due to their simplicity. For example, in the field of machine learning research, script-based development tools such as Jupyter Notebooks are widely used \cite{prathanrat2018jupyternotebook}. Similarly, multiple approaches have been proposed to reduce the overhead of programming language expertise in Cloud environment simulation using script-based \cite{tsakanikas2022vfcsim, jammal2019gits, silva2014cloudsimplusautomation} or graphical user interface (GUI)-based \cite{wickremasinghe2012cloudanalyst, nunez2012icancloud, toledo2020epcsac, TeixeiraSá2014cloudreports, li2012dartcsim, jammal2019gits} Cloud environment modeling. In doing so, both script-based and GUI-based approaches compromise on the extensibility of the simulation platform. For example, Silva et al. \cite{silva2014cloudsimplusautomation} propose a human-readable script to define the Cloud environment, which is then consumed by a specific software tool to implement the simulation scenario. However, the script is limited to a specific set of allocation policies that can be used. Similarly, DartCSim \cite{li2012dartcsim} attempts to customize the simulation platform by providing an embedded code editor, but only a limited set of methods are allowed to be overridden. Compromised extensibility reduces the flexibility in creating simulation scenarios. This motivates us to investigate reducing programming language overhead in simulating Cloud computing environments, with a minimum compromise in simulation platform extensibility.

\par In this paper, we propose an architectural framework to translate a script-based Cloud computing environment into an actual simulation in a simulation platform. Our framework segregates the simulation platform and translation logic into different layers, preserving the extensibility of the simulation platform while providing a simplified script-based simulation. To validate the proposed framework, we implement it for the widely used Cloud environment simulator, the CloudSim toolkit \cite{calheiros2011cloudsim}. The resulting framework, called \textit{CloudSim Express}, enables human-readable script-based Cloud environment simulations. It consumes a YAML (YAML Ain’t Markup Language) \cite{yaml} script that describes the simulation system model in a top-down approach and automates code-based simulation scenario implementation. It separates the simulation platform extension logic by allowing dynamic injection of extensible modules via the YAML script. Furthermore, it utilizes the layered architecture of the proposed framework to configure extensible modules with the CloudSim toolkit.

We evaluate the performance of\textit{CloudSim Express} against the state-of-the-art baselines using a practical use case of Cloud environment simulation. Unlike the baselines, our \textit{CloudSim Express} framework makes minimal compromise in simulation platform extensibility while significantly reducing the overhead of programming language expertise. \textit{CloudSim Express} outperforms the baselines with a reduction of over $71.43$\% in code complexity and over $89.42$\% in lines of code. Additionally, for the purpose of reproducibility and broader use by the research community, the developed solution is released as an open-source project. In summary, our contributions are,
\begin{itemize}
    \item We propose an architectural framework to translate an script-based Cloud environment system model, into a concrete simulation
    \item We validate the proposed architectural framework by implementing it for the widely-adopted CloudSim toolkit \cite{calheiros2011cloudsim}
    \item We release the implemented framework, the \textit{CloudSim Express}, as an open source framework
\end{itemize}

The rest of the paper is organized as follows. Section \ref{section:background} discusses related background, Section \ref{section-archi-framework} discusses our proposed Architectural framework, Section \ref{section-cloudsim-express} discusses our implementation of the proposed framework, Section \ref{section-evaluation} discusses the comparison and the evaluation of the implementation, Section \ref{section-related-work} discusses the related literature, and Section \ref{section-conclusion-future-direction} concludes our work with potential future work.

\section{Background}
\label{section:background}

\subsection{Cloud Environment and Simulation}

A Cloud environment is a large and complex system composed of users, computing resources, network facilities, and applications. It focuses on various characteristics, including geographic location awareness, low latency, application scalability, and elasticity. As it evolves, new challenges emerge, such as energy efficiency and load balancing \cite{mansouri2020cloussimulatorsurvey}. However, conducting experiments in actual Cloud environments is expensive, time-consuming, and difficult to replicate. Furthermore, developers lack control over Cloud environments, making it infeasible to repeat benchmark scenarios \cite{mansouri2020cloussimulatorsurvey}.

Simulation allows for the imitation of real-world Cloud environments and enables experimentation with their complex internal interactions \cite{mastenbroek2021opendc}. It enables the stress-testing and iterative improvement of new strategies through repeated benchmark scenarios. Moreover, simulation requires fewer resources and less time to conduct experiments. Therefore, the use of Cloud environment simulators for experimenting with new strategies prior to actual deployment is common in research \cite{li2012dartcsim}.

\subsection{Extensible Cloud Simulators}

Cloud simulators provide a system model to represent a cloud environment. For example, the CloudSim toolkit follow a system model consisting of cloud brokers and data centers \cite{calheiros2011cloudsim}. Another example is the GreenCloud simulator, which focuses on network and energy and has a system model consisting of data centers and network devices with their energy models \cite{kliazovich2012greencloud}. In most simulators, the system models are extensible, meaning they provide extension points to inject scenario-specific logic. Users can leverage these extension points to experiment with novel approaches, such as scheduling policies and allocation policies \cite{mansouri2020cloussimulatorsurvey}.

Furthermore, the simulation platform itself can be extensible. This is the case for widely used simulators like the CloudSim toolkit, which allows for reusable code \cite{calheiros2011cloudsim}, and enables the creation of extended simulators that are use case specific \cite{mansouri2020cloussimulatorsurvey}. For example, CloudSimSDN provides support for Software Defined Networks in CloudSim by extending the vanilla CloudSim toolkit \cite{son2015cloudsimsdn}. Another example is GreenCloud, which extends the Network Simulator NS2 \cite{issariyakul2009ns2} and adds cloud environment components to create an energy and network-focused simulator \cite{kliazovich2012greencloud}.

Therefore, the extensibility of cloud simulators, which involves providing extension points and the ability to extend itself, is significant in accommodating various cloud simulation scenarios. Most simulators leverage programming language features, such as object-oriented representation of components, to implement extensibility \cite{cloudsimplus, calheiros2011cloudsim}. However, these implementations often come with a steep learning curve. At the same time, simplifying the cloud environment system models (e.g., using script-based system models) to reduce the learning curve can compromise extensibility, as they encapsulate the object-oriented representation of inner components \cite{li2012dartcsim, wickremasinghe2012cloudanalyst}.

\subsection{CloudSim Toolkit}

The CloudSim toolkit is a Java-based event-driven simulation toolkit for Cloud environments \cite{calheiros2011cloudsim}. It is widely used in research \cite{silva2014cloudsimplusautomation, jammal2019gits} due to its rich features and its ability to be extended and customized to cater to most simulation scenarios. 
{%
CloudSim supports the virtualization of physical machines, offering customizable provisioning policies for Virtual Machines (VMs). This capability extends to modeling and simulating application containers. Moreover, CloudSim's power packages facilitate the modeling of the energy aspects of computations, thereby enabling the simulation of energy-optimization techniques. CloudSim evolves as the field advances enabling researchers to experiment with novel concepts. Such examples features are Microservices architectures, Edge/Fog computing \cite{ifogsim22022redowan}, Serveless computing \cite{mampage2023cloudsimsc}, and Quantum computing \cite{nguyen2023iquantum}. The architecture of CloudSim allows for dynamic insertion of components, enhancing the management of the simulation lifecycle with features like stop-and-resume. The simulation environment's interconnected networks can be effectively modeled using network topologies. This setup allows for an analysis of the impact of network characteristics, such as latency and bandwidth, on the communication between components. Furthermore, CloudSim includes user-defined policies for resource management, which can be tailored to various customized scenarios, such as allocating VMs to host resources or assigning processing elements to active VMs in a granular manner. This level of customization reduces the effort required for system model development, enabling researchers to concentrate more on aspects of resource management.}

{%
\par The CloudSim toolkit has garnered significant attention within the scientific community, continuously fostering a rich ecosystem of extensions. One notable extension is WorkflowSim, developed by the University of Southern California, which adds a higher layer of workflow management to CloudSim. This layer addresses heterogeneous system overheads and failures \cite{workflowsim2012chen}, enabling effective evaluations of scientific workflows in distributed settings. A distinctive aspect of modern cloud infrastructures is the investigation of cloud storage and its energy consumption. In this context, CloudSimDisk \cite{cloudsimdisk2015louis} proposes a scalable module for CloudSim, extending its capabilities to support energy-aware storage solutions. This module has been validated by comparing simulation results with analytical models designed to gauge the energy consumption of hard disk drives in cloud systems. NetworkCloudSim addresses the lack of advanced application models in heterogeneous clouds. It focuses on scalable networking and a generalized application model, encompassing applications such as message passing and workflows. This allows for more accurate evaluations of scheduling and resource provisioning policies \cite{networkcloudsim2011garg}. The migration of software systems to the cloud necessitates the evaluation of competing cloud deployment options (CDOs). To this end, CDOSim offers a tool for assessing the cost and performance properties of CDOs \cite{cdosim2021fittkau}, integrating a migration framework with CloudSim and leveraging provider-specific performance characteristics.

Furthermore, CloudSimEx aims to maintain a collection of extensions that are not only research-oriented but also focused on improving software engineering aspects. These improvements include enhanced logging utilities, the generation of CSV files for analysis purposes, and the capability to run multiple experiments in parallel \cite{cloudsimex}. Cloud2Sim concentrates on executing CloudSim simulations in a distributed manner, utilizing resources such as multi-core CPUs \cite{cloud2sim2014kathiravelu}. It achieves this through a distributed object storage approach. CloudSimSDN \cite{son2015cloudsimsdn} introduces software-defined networking (SDN) modeling to CloudSim simulations, enabling researchers to explore SDN-oriented resource management approaches. The diverse ecosystem of CloudSim extensions highlights CloudSim's ability to offer a versatile simulation framework for cloud computing.
}

At the same time, the CloudSim toolkit, and {most of its extensions are based on modules that are written in java}, which means that its simulations are developed as Java projects. Therefore, an additional overhead of programming language expertise is added to the development life cycle. Besides, CloudSim does not specify a standard pattern for implementing simulation scenarios, making it complex to reuse common components across simulations. This also leads to simulation code bases that are difficult to manage and have poor code readability. Aforementioned drawbacks are primarily caused by the lack of a standard framework for developing modularized simulations and the absence of a simplified representation of the simulation system model, such as a human-readable script. A modularized framework would enable CloudSim users to share common simulation components, drastically improving code readability and maintainability. A simplified representation of the simulation system model would allow the development of simplified simulations using a top-down approach without requiring expertise in a programming language. 

{%
To that end, existing literature aim to minimize the programming language overhead in CloudSim simulations. These works focus on providing a Graphical User Interface (GUI) to facilitate the design and execution of simulations through CloudSim \cite{wickremasinghe2012cloudanalyst, li2012dartcsim}. They enable users to set configurable parameters via the GUI and observe statistics of the ongoing simulation. The underlying implementation is responsible for translating between the GUI and CloudSim. A similar approach can be implemented using a human-readable script \cite{silva2014cloudsimplusautomation}. This method eliminates the need for users to have proficiency in a programming language, as the script and its structure for modeling system components are designed to be human-readable.}

{%
However, while approaches aimed at reducing programming overhead can streamline the use of CloudSim, they may also limit the toolkit's extensibility. Extensibility is an important feature of CloudSim, forming the foundation of its extensive ecosystem of extensions. Compromising on this aspect diminishes the ease with which implemented approaches can be integrated into existing extensions. For instance, DartCSim, which offers a GUI interface, permits the overriding of only a limited set of methods \cite{silva2014cloudsimplusautomation}, thus constraining the scope for customization. Similarly, CloudSimPlus Automation, which employs a human-readable script to reduce programming complexity, restricts users in customizing implementations due to its specific usage of the Datacenter component \cite{silva2014cloudsimplusautomation}. Addressing this gap, our work introduces a novel framework that significantly reduces programming overhead while minimally impacting CloudSim's extensibility.
}

\section{Architectural Framework}
\label{section-archi-framework}

In this section, we propose a generic modularized architectural framework for designing the Cloud environment simulation system model using a human-readable script that can be automatically translated into a concrete simulation implementation.

\begin{figure*}[t]
\centerline{\includegraphics[scale=0.6]{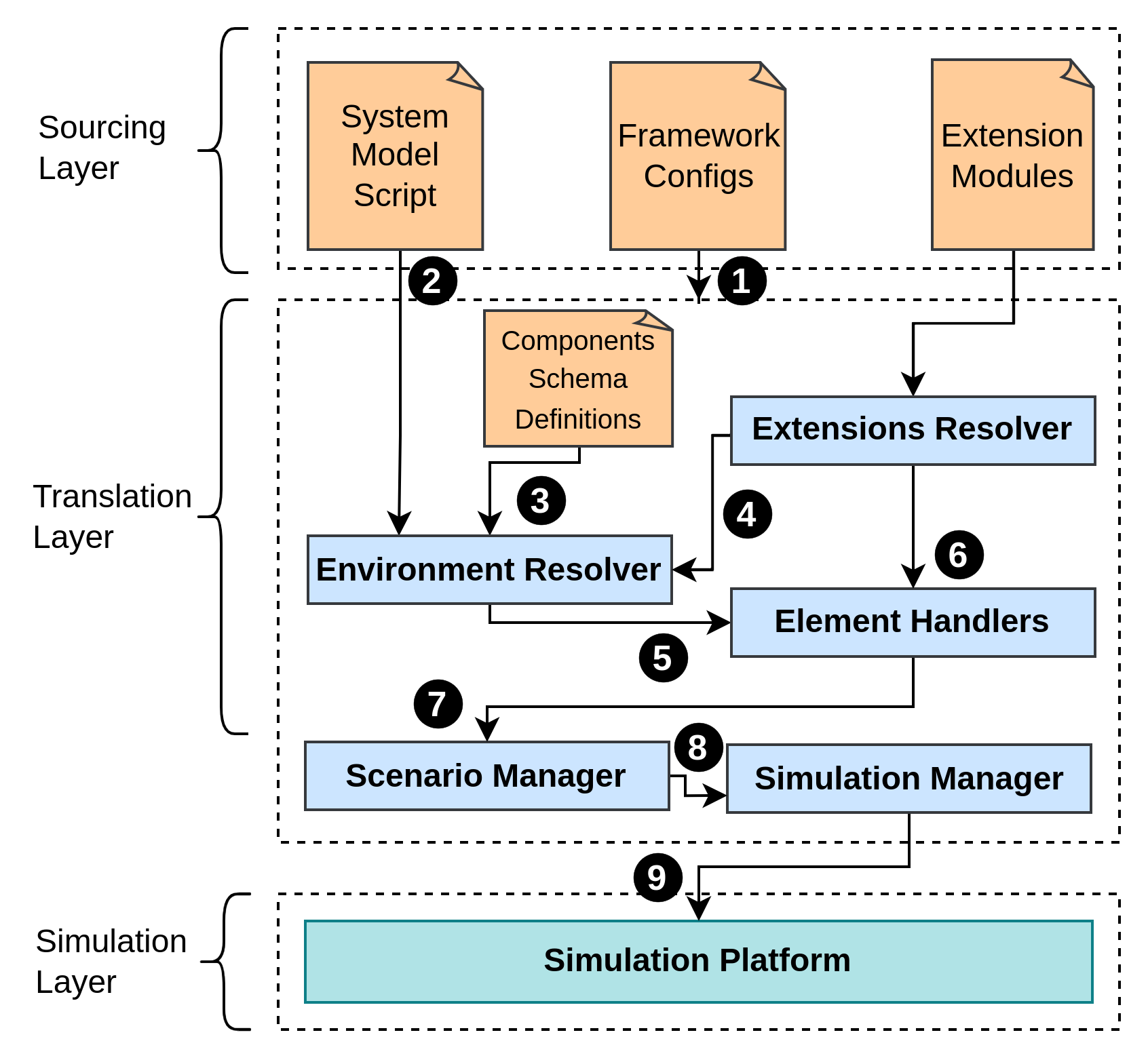}}
\caption{Architectural Framework for Script-based Simulations\label{fig:low-code-framework}}
\end{figure*}

Figure \ref{fig:low-code-framework} illustrates the proposed architectural framework. Our architecture consists of three main layers: the \textit{Sourcing Layer}, which maintains user-modifiable files and configurations; the \textit{Translation Layer}, which handles the logic of translating the \textit{System Model Script} into an actual simulation; and the \textit{Simulation Layer}, which represents the simulation platform. The following sections provide a detailed description of each layer and explain the overall control flow for translating the \textit{System Model Script}  into a concrete simulation.
\subsection{Sourcing Layer}

The \textit{Sourcing Layer} provides the following external sources.

\begin{itemize}
    \item\textbf{System Model Script:} The simulation system model is described using a human-readable scripting language. The system model comprises multiple interconnected components and can be simplified using a top-down approach. For instance, a regional Cloud zone itself is a system model that includes one or more Datacenters. Each Datacenter consists of a set of hosts, and these hosts, in turn, have associated processing elements. Thus, the \textit{System Model Script} describes each of these components and aggregates them into an overall system model component.
    \item\textbf{Framework Configurations:} Provides various configurations required to initialize the framework, such as location information of the files in the \textit{Sourcing Layer}.
    \item\textbf{Extensions Modules:} Provides modularized extensions to the user. These extensions are used to customize the simulation platform, as well as the framework itself.
\end{itemize}

\subsection{Translation Layer}

This layer consists of the following modularized components to translate the human-readable system model into a concrete simulation.

\begin{itemize}
    \item\textbf{Environment Resolver:} Parses the \textit{System Model Script} file, into a set of components that describes the system model. It uses the supplied component schema to understand the \textit{System Model Script}. Upon parsing, components are aggregated into an overall system model component and handed over to the corresponding \textit{Element Handler}. For example, a host described in the \textit{System Model Script} can contain $7$ processing elements. This is parsed into a host component having its processing elements attribute set to $7$. Afterwards, multiple hosts components are aggregated to create the Datacenter system model, which is then handled by a corresponding \textit{Element Handler}.
    
    \item \textbf{Element Handler:} Processes a component in the simulation environment, and provides the corresponding simulation logic that needs to be implemented. Therefore, an \textit{Element Handler} is implemented per Simulation Platform, and injected into the \textit{Translation Layer} as an extension.

    \item \textbf{Extensions Resolver:} Provides a central resolver to materialize objects from the provided extension modules. It generates extensions for various \textit{Element Handlers}, \textit{Translation Layer} modules, and the extensible components in the \textit{Simulation Platform}.

    \item \textbf{Scenario Manager:} Handles simulation scenario by executing \textit{Element Handler} components.

    \item \textbf{Simulation Manager:} Manages the overall simulation, such as  initializing the simulation platform, initiating the system model translation, starting the simulation and overseeing it's execution.
    
\end{itemize}

\subsection{Simulation Layer}

This layer consist the simulation platform. The \textit{Translation Layer} interacts with it during the initialization and simulation execution stages. It  decouples the simulation platform logic from the script-based translation logic, thereby allowing framework to be implemented with different simulation platforms. 

\subsection{System Model Script to Simulation Translation flow}

Fig. \ref{fig:low-code-framework} denotes the following control flow of the architectural framework.

\begin{enumerate}
    \item \ballnumber{1} initialize the \textit{Translation Layer}. This includes registering available \textit{Element Handlers}, initializing \textit{Extensions Resolver} with the location of extension modules, etc.
    
    \item \ballnumber{2} and \ballnumber{3} denotes \textit{Environment Resolver} parsing the system model. It reads the \textit{System Model Script} and constructs the system model component using the schematics of the components provided with \ballnumber{3}. Upon construction of the system model component, its associated \textit{Element Handler} is materialized via the \textit{Extensions Resolver} in \ballnumber{4}, and then parsed system model component is handed over.
    
    \item \ballnumber{5}, \ballnumber{6}, and \ballnumber{7} denotes \textit{Scenario Manager} handling the simulation scenario using the system model element handler. In doing so, it materializes other associate \textit{Element Handlers} of the system model element handler in a top-down approach.
    
    \item \ballnumber{8} and \ballnumber{9} denotes \textit{Simulation Manager} managing the simulation by executing the simulation scenario from \ballnumber{8}, with the \textit{Simulation Platform} provided by \ballnumber{9}.
\end{enumerate}

\section{Implementation}
\label{section-cloudsim-express}

In this section, we describe the implementation of our proposed framework for CloudSim toolkit.

\subsection{System Model Script and Components Schema Definitions}
\label{section:schema-to-pojo}

To simplify the Cloud environment system model with a human-readable script, we choose the human-friendly data serialization language, YAML (YAML Ain't Markup Language) \cite{yaml}. YAML provides the capability to describe multiple components and then to aggregate the system model. Besides, YAML is widely used in the industry from managing configurations \cite{lei2022yamlconfigs}, to describing APIs, thereby provides a familiar experience to users.

To define the schema to write components in the \textit{System Model Script} , we use OpenAPI 3.0 specification \cite{openapi3}. OpenAPI 3.0 is a machine-readable interface definition language and it is widely used with web services \cite{romero2023openapi}. It provides specifications to describe object schematics in a top-down approach. 

\noindent\begin{minipage}{.45\textwidth}
\begin{lstlisting}[caption={Component Schematics},label={listing:compile-schematics}, basicstyle=\fontsize{8}{10}\selectfont\ttfamily]
Datacenter:
    type: object
    properties:
        variant:
          $ref: '#/components/schemas/Extension'
    characteristics:
        $ref: '#/components/schemas/Datace...'
    vmAllocationPolicy:
    ...
\end{lstlisting}
\end{minipage}\hfill
\begin{minipage}{.45\textwidth}
\begin{lstlisting}[caption={Generated Java Class},label={listing:compile-generated-class}, basicstyle={\fontsize{8}{10}\selectfont\ttfamily}]
@Generated(value = "org.openapi...")
public class Datacenter {

  @JsonProperty("variant")
  private Extension variant;

  @JsonProperty("characteristics")
  private DatacenterCharacteristics cha...
...
\end{lstlisting}
\end{minipage}

\noindent\begin{minipage}{.45\textwidth}
\begin{lstlisting}[caption={System Model Script},label=runtime-yaml, basicstyle=\fontsize{8}{10}\selectfont\ttfamily]
Datacenter: *Datacenter
  variant:
    className: "org.cloudbus.cloudsim.Datacenter"
  characteristics: *Characteristics
  vmAllocationPolicy:
    className: "org.cloudbus.cloudsim.VmAllocation..."
  storage: ""
  schedulingInterval: 0
...
\end{lstlisting}
\end{minipage}\hfill
\begin{minipage}{.45\textwidth}
    \captionof{figure}{Runtime Parsed Datacenter Object}
    \includegraphics[scale=0.19]{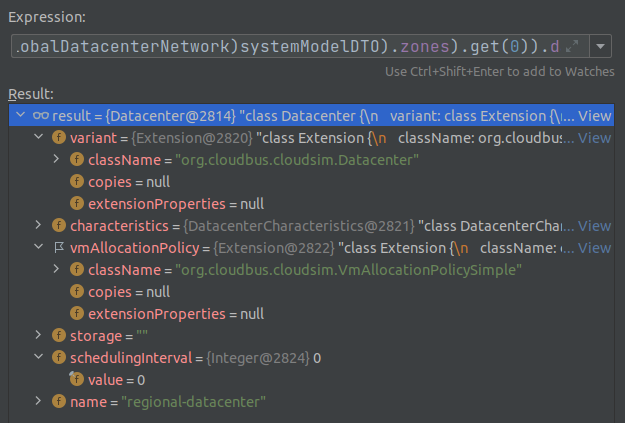}
    \label{fig:runtime-parsed-object}
\end{minipage}

\hspace{1cm}

\par Firstly, we use OpenAPI code generation library during \textit{CloudSim Express} compilation time to read components schema definitions and convert them to plain-old-java-object (PoJo) classes. During its run time, \textit{CloudSim Express} reads the \textit{System Model Script} YAML file and parse YAML components to objects of aforementioned PoJo classes. Both compilation time and run time translations of the Datacenter component are demonstrated with Listing \ref{listing:compile-schematics}, \ref{listing:compile-generated-class}, \ref{runtime-yaml} and Fig. \ref{fig:runtime-parsed-object}. In which, Listing \ref{listing:compile-schematics} shows the object schematics of the Datacenter component, alongside the corresponding attribute names and the type of the value complying with OpenAPI 3.0 specification. Listing \ref{listing:compile-generated-class} shows the PoJo class generated upon compilation of the \textit{CloudSim Express}. Listing \ref{runtime-yaml} shows the datacenter information sourced from the \textit{System Model Script}, for a given simulation scenario. This information is written according to the schematics defined in Listing \ref{listing:compile-schematics}. As shown in Fig. \ref{fig:runtime-parsed-object}, this information is then translated into a PoJo object during \textit{CloudSim Express} run time.

\subsection{Environment Resolver }
\label{section:env-resolver-impl}

The Section \ref{section:schema-to-pojo} describes how \textit{CloudSim Express} translates \textit{System Model Script} to a system model PoJo object. The run time translation process of that (also demonstrated with Listing \ref{runtime-yaml} and Fig. \ref{fig:runtime-parsed-object} is implemented with the \textit{Environment Resolver} component. It uses a YAML parsing library to read the \textit{System Model Script} and map the parsed information into the PoJo object using an object mapping library. Afterwards, it traverses through registered Element Handlers and materializes a new matching Element Handler for the system model PoJo object.

\subsection{Extensions Resolver }

An extension in \textit{CloudSim Express} is a java module, packed as a jar. Extensions can extend the CloudSim toolkit, as well as the \textit{CloudSim Express} framework itself. The \textit{Extensions Resolver} module implementation provides APIs to materialize extension objects. During the launch of \textit{CloudSim Express}, it reads and loads the class files packed in extension jar modules. It then provides APIs to allow other modules in the \textit{Translation Layer} to invoke extension creation requests, with necessary information provided as custom java constructor arguments. We use java reflections feature to create objects using the loaded extension classes and the provided constructor arguments.

\subsection{Segregating CloudSim Simulation logic into Element Handlers}
\label{section-seggregation}

\begin{figure}[h]
    \centering
    \includegraphics[scale=0.6]{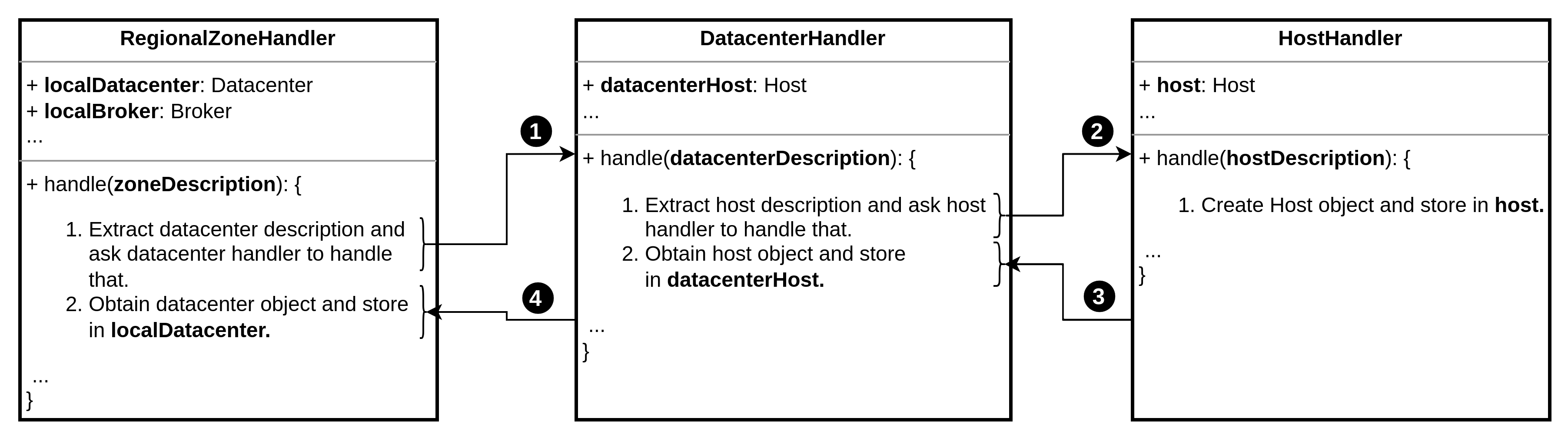}
    \caption{Segregation of CloudSim Simulation Logic with Element Handlers}
    \label{fig:seggregation-element-handlers}
\end{figure}

A standard CloudSim simulation involves building the system model in a java method body. We segregate this implementation to a series of recursive calls of Element Handler components. Each Element Handler provides implementation logic to a specific component in its \textit{handle()} method, and for any other associated components, it makes a recursive method call of the same method of the corresponding Element Handler. This flow is demonstrated with Fig. \ref{fig:seggregation-element-handlers}. In which, a regional zone handler require a datacenter object for its implementation logic. It then calls \textit{handle()} method of the datacenter handler and obtain the required datacenter object. The datacenter handler requires a host object for its implementation logic, which then is obtained by calling the same method of the host handler. The actual control flow for this scenario is depicted by the number \ballnumber{1} to \ballnumber{4}.
\\
Overall, the \textit{handle()} method instantiates the required CloudSim classes and creates the elements that are required. A series of element handler calls then generate all the required instances of CloudSim classes to establish the simulation environment. These element handlers can be easily customized to meet user requirements through the schema definition file. In this file, users define the schema for new elements and update the framework with corresponding POJO (Plain Old Java Object) files, which is automated during the build process. Subsequently, the corresponding element handler implementations are dynamically injected into the framework.

\subsection{Execution of the Simulation}

Once the Element Handler of the system model is realized (i.e. the process discussed in Section \ref{section:env-resolver-impl}), it is then passed to the Scenario Manager. The Scenario Manager is controlled by the Simulation Manager. When \textit{CloudSim Express} starts executing, the Simulation Manager instructs Scenario Manager to build the simulation scenario. The Simulation Manager achieves this by calling the \textit{handle()} method of the system model Element Handler. As explained in Section \ref{section-seggregation}, this eventually implements the CloudSim logic for building the system model. Afterwards, the Simulation Manager starts the simulation with CloudSim platform and oversee towards its completion.

\subsection{Managing Extensions}

Being able to extend is a powerful feature of CloudSim. Our proposed platform provide the flexibility to keep that intact, by integrating CloudSim extensions with \textit{Extensions Resolver}. We define an attribute \textit{variant}, in the \textit{System Model Script} to indicate that a custom version of a component needs to be used. Listing \ref{runtime-yaml} and Fig. \ref{fig:runtime-parsed-object} demonstrate an example of this. Which, the datacenter component that needs to be used is of the type \textit{org.cloudbus.cloudsim.Datacenter}. This block of information is passed to the relevant element handler, and the element handler materializes a datacenter component of the mentioned type via the \textit{Extensions Resolver}, through the java reflection feature. The user needs to place the extension class packed into a jar in the extensions location, and set the corresponding class name in the \textit{System Model Script} file. Afterwards, the \textit{CloudSim Express} tool is able to interpret the relevant extended version.

Apart from the simulation platform, the components in the \textit{CloudSim Express} framework can also be extended in the same manner. However, since this is not a part of the simulation, the user is required to place the corresponding jar same as before, but the corresponding class name needs to be configured in the configuration file of \textit{CloudSim Express}. This feature is useful if a user wants to change the default behaviour of a component in the \textit{Translation Layer}, such as changing the behaviour of a default Element Handler.

\subsection{Generalizability}

Apart from the \textit{CloudSim Express} implementation, the proposed framework is general enough to be applicable to any other simulator that builds the simulation system model using a top-down approach. In such scenarios, the implementation logic specific to the simulator is customized in the \textit{Element Handler} components, and the \textit{Translation Layer} components are customized as needed. Each of these components is designed to be pluggable, making the deployment of customized components straightforward.

\section{Performance Evaluation}
\label{section-evaluation}

To evaluate the effectiveness of our framework, we implement a use case with \textit{CloudSim Express} and baselines. We collect qualitative metrics and quantitative aspects of the development life cycle productivity.

\subsection{Design of the Use Case}
\label{section-use-case-design}

CloudSim simulations can range from experimenting with allocation policies to experimenting with novel infrastructure changes, such as evaluating low power datacenters. These scenarios are implemented by extending the CloudSim toolkit. Therefore, in order to evaluate true productivity improvements from \textit{CloudSim Express}, the use case should stress extensibility of the CloudSim toolkit.

The standard extension points of the CloudSim toolkit support a variety of policies, such as virtual machine allocation. In addition to that, it can be further extended using its object-oriented architecture. To capture both extensible aspects, we design following use case requirements.

\begin{itemize}
    \item \textbf{R1:} A datacenter that is aware of its Virtual Machines (VMs) allocation to physical hosts.
    \item \textbf{R2:} A datacenter that periodically monitors its Virtual Machines (VMs)
\end{itemize}

The CloudSim toolkit utilizes the abstract Java class \textit{VmAllocationPolicy} to define the logic for allocating VMs to physical machines during the simulation. By default, CloudSim includes an implementation of the worst-fit algorithm (i.e., allocating a VM to the host with the most processing elements) through the concrete class \textit{VmAllocationPolicySimple}, which extends the aforementioned abstract class. In practical scenarios, users can implement their own VM allocation algorithms by extending the same abstract class. Therefore, \textbf{R1} is also implemented by creating a new class that extends the \textit{VmAllocationPolicy} abstract class. Once the implementation is completed, the CloudSim simulation code needs to be modified to include the newly created allocation policy class, and the codebase must be recompiled before running the simulation.

The CloudSim extension points do not cover all stages of the simulations. Anything that is not covered by the standard simulation points is implemented through the object-oriented design of the CloudSim toolkit. In such scenarios, users extend specific classes that represent the components in the simulation system model. Therefore, \textbf{R2} is implemented by extending the standard CloudSim class "Datacenter" and overriding the method "updateCloudletProcessing". Within the overridden method, the logic corresponding to periodic virtual machine monitoring is implemented. Afterward, the extended "Datacenter" class needs to be included in the CloudSim simulation code, and the codebase must be recompiled before running the simulation.

In both implementations, users are required to be familiar with the CloudSim simulation codebase and need to invest effort into recompiling the codebase multiple times. Overall, this process involves significant effort. However, it offers users fine-grained flexibility in designing the simulation. Therefore, an ideal approach for reducing the aforementioned programming language overhead should also involve minimal compromise in simulation design flexibility.

\subsection{Baseline Approaches}

We use the following baseline approaches that reduces the programming language overhead of the CloudSim toolkit.

\begin{itemize}
    \item \textbf{CloudSim Plus Automation \cite{cloudsimplusautomationgithub,silva2014cloudsimplusautomation}:} A YAML (YAML Ain't a Markup Language) script-based simulation approach for CloudSim simulations.
    \item \textbf{GITS: Generic Input Template for Cloud Simulators \cite{jammal2019gits}:} A JSON (JavaScript Object Notation) script-based generic input template based approach for CloudSim simulations. 
\end{itemize}

\subsection{Implementation with \textbf{CloudSim Plus Automation}}

The CloudSim Plus Automation provides YAML script-based system model design. We implemented the use case by designing a single Datacenter, with a single customer. However, the CloudSim Plus Automation does not have native flexibility for custom extensions. It only supports the standard extension points supported with the CloudSim Plus project \cite{silva2017cloudsimplus}, which is an enhanced version of the CloudSim toolkit. Therefore, to implement \textbf{R1} and \textbf{R2} from Section \ref{section-use-case-design}, we follow the approach below after inspecting the source code of CloudSim Plus Automation.

\begin{itemize}
    \item \textbf{R1:} Write an extended VM allocation policy class, \textit{VmAllocationPolicyCustom}, using the java package, \textit{org.cloudsimplus.allocationpolicies}. This is because CloudSim Plus Automation only supports allocation policy classes with the naming prefix \textit{VmAllocationPolicy}, and residing in the aforementioned java package. When a custom class is implemented in this manner, it can be selected via the YAML script.
    \item \textbf{R2:} Write an extended \textit{DatacenterSimple} class. The \textit{DatacenterSimple} class represents datacenters in CloudSim Plus Automation. In order to select it, we modified the source code. Because selecting a custom datacenter class is not supported via the YAML script.
\end{itemize}

\noindent\begin{minipage}{.33\textwidth}
\begin{lstlisting}[caption={CloudSim Plus Automation: \\ YAML Script},label=listing:system-model-script:cloudsim-plus-automation, basicstyle=\fontsize{8}{10}\selectfont\ttfamily]
datacenters:
  - !datacenter
    amount: 1
    vmAllocationPolicy: Custom
    ...
    sans:
      - !san
        ...
    hosts:
      - !host
        amount: 5
        ...
\end{lstlisting}
\end{minipage}\hfill
\begin{minipage}{.33\textwidth}
\begin{lstlisting}[caption={GITS: JSON Input\\ Template},label=listing:system-model-script:gits, basicstyle=\fontsize{8}{10}\selectfont\ttfamily]
{
  "name": "GITS template",
    ...

  "DC": {
    "name": "RegionalDC",
    ...

  "Rack": {
    ...

    ...
\end{lstlisting}
\end{minipage}
\begin{minipage}{.33\textwidth}
\begin{lstlisting}[caption={CloudSim Express:\\ System Model Script},label=listing:system-model-script:cloudsim-express, basicstyle=\fontsize{8}{10}\selectfont\ttfamily]
- <<: *Host
    ...
      copies: 5
Characteristics: &Characteristics
  ...
Datacenter: &Datacenter
  variant:
    className: "org.cloudbus...
  characteristics: *Character...
  vmAllocationPolicy:
    className: "org.cloudbus...
  ...
\end{lstlisting}
\end{minipage}

To load custom classes during runtime, both were implemented in the CloudSim Plus Automation project. Afterward, CloudSim Plus Automation project was recompiled to obtain a modified tool with custom classes. Finally, the tool was executed with the corresponding YAML script, which is denoted in Listing \ref{listing:system-model-script:cloudsim-plus-automation}. In this script, we set VM allocation policy value as \textit{Custom}, so that our custom allocation policy implementation is selected.

\subsection{Implementation with \textbf{GITS (Generic Input Template for Cloud Simulators)}}
\label{section-gits-impl}

The GITS (Generic Input Template for Cloud Simulators) focuses on representing the Cloud environment with either textual (JSON (JavaScript Object Notation) template) or graphical representation, and for this comparison, we use the former. The GITS for CloudSim provides a JSON template describing the Cloud Environment, and a GITS library file to convert Datacenter and hosts information from the JSON template to the GITS java objects.

\begin{figure*}[h!]
\centerline{\includegraphics[scale=0.6]{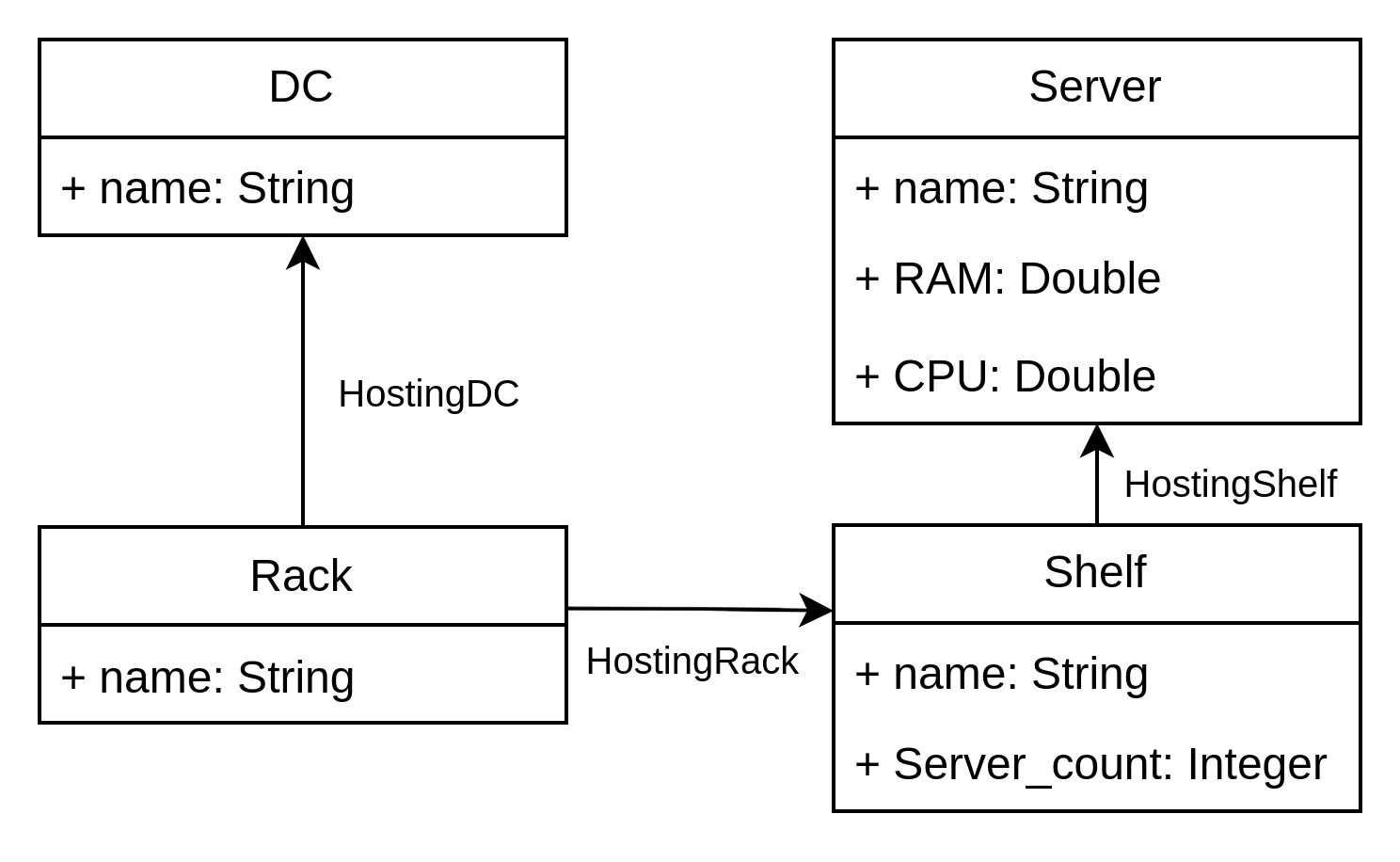}}
\caption{GITS UML Diagram for Generic Input Template
    \label{fig:gits-uml}}
\end{figure*}

The Fig. \ref{fig:gits-uml} depicts the UML diagram for the objects in the JSON template corresponds to the use case. We implemented GITS library file with the logic to handle objects in the template \cite{jammal2019gits}. Once GITS java objects are created, the CloudSim Datacenters and hosts are populated based on their values. Therefore, we implemented the use case requirements with the standard CloudSim simulation described in Section \ref{section-use-case-design} while using GITS as a layer to interpret datacenter and host values from the JSON input template, which is denoted in Listing \ref{listing:system-model-script:gits}.

\subsection{Implementation with \textbf{CloudSim Express}}

The system model of \textit{CloudSim Express} implementation includes a regional zone with a datacenter. The extensions required by \textbf{R1} and \textbf{R2} from Section \ref{section-use-case-design} are implemented as follows.

\begin{itemize}
    \item \textbf{R1:} Write an extended VM allocation policy class
    \item \textbf{R2:} Write an extended \textit{Datacenter} class
\end{itemize}

Both custom classes were developed in a separate java project, and compiled into a jar file. The resulting jar file is then copied to the \textit{CloudSim Express} tool. 

Listing \ref{listing:system-model-script:cloudsim-express} denotes \textit{System Model Script}  for the use case, in which we modified it to select the two custom classes. 
{
This \textit{System Model Script}  is also copied to the tool. Since default element handlers follow the commonly used cloud architectures, we do not need to modifiy them. Therefore, we executed \textit{CloudSim Express} tool, which reads the script and load the custom classes during starup, and then to execute the simulation. \textit{CloudSim Express} tool require least effort in deploying custom simulation scenarios, as the tool has automated most of the class loading procedures.

\subsection{Productivity Analysis}

Productivity can be measured in two aspects: quantitative and qualitative. We measure the quantitative aspect by considering the amount of programming involved. The primary goal of a scripted simulation is to minimize the time spent on coding. Table \ref{perf-compare-table} depicts the differences in productivity between \textit{CloudSim Express} and the baseline approaches for the implemented use case.

We measure the lines of code to gauge the project's magnitude. Additionally, we measure the complexity of the code because customizing the use case sometimes requires analyzing and modifying the simulation platform code. If the code is more complex, it takes more time to analyze. Since all approaches are Java-based solutions, we utilize JaCoCo, an open-source code coverage report generation tool \cite{jaCoCo-counters}, to measure the lines of code and code complexity. JaCoCo calculates cyclomatic complexity value for each non-abstract method in the code, which is an indication of number of unit tests required for possible paths through the methods. A higher complexity number require increased effort in analyzing the control flow of the code. For both measurements, we excluded the GITS library file implementation code from Section \ref{section-gits-impl} as it can be imported as a library.

We measure four qualitative aspects: framework re-compilation, CloudSim implementation, runtime extension injecting, and extending via script support. Framework re-compilation indicates whether modifying the provided framework is required to achieve the customized behavior. CloudSim implementation indicates whether at least a partial CloudSim simulation code implementation is needed. Runtime extension injecting indicates whether extended components are dynamically provided without re-compiling the framework. Extending via script support indicates whether framework extendability is achieved via the \textit{System Model Script} .

\begin{table}[t]
\resizebox{\textwidth}{!}{
    \begin{tabular}{l|ll|llll|}
    \cline{2-7}
                                                   & \multicolumn{2}{c|}{\textbf{Quantitative}}                                                                                                                                 & \multicolumn{4}{c|}{\textbf{Qualitative}}                                                                                                                                                                                                                                                                                                                                                                                   \\ \hline
    \multicolumn{1}{|l|}{\textbf{Approach}}        & \multicolumn{1}{l|}{\textbf{\begin{tabular}[c]{@{}l@{}}JaCoCo:\\ Code Complexity\end{tabular}}} & \textbf{\begin{tabular}[c]{@{}l@{}}JaCoCo:\\ Lines of Code\end{tabular}} & \multicolumn{1}{l|}{\textbf{\begin{tabular}[c]{@{}l@{}}No Framework \\ Re-compilation\end{tabular}}} & \multicolumn{1}{l|}{\textbf{\begin{tabular}[c]{@{}l@{}}No CloudSim \\ Implementation\end{tabular}}} & \multicolumn{1}{l|}{\textbf{\begin{tabular}[c]{@{}l@{}}Runtime Extension \\Injecting\end{tabular}}} & \textbf{\begin{tabular}[c]{@{}l@{}}Extending \\ via Script\end{tabular}} \\ \hline
    \multicolumn{1}{|l|}{\text{CloudSim Plus Automation}} & \multicolumn{1}{l|}{314}                                                                         & 740                                                                      & \multicolumn{1}{l|}{{\color[HTML]{FE0000} \xmark}}                                                  & \multicolumn{1}{l|}{{\color[HTML]{009901} \cmark}}                                                  & \multicolumn{1}{l|}{{\color[HTML]{FE0000} \xmark}}                                                        & {\color[HTML]{009901} \cmark}                                                 \\ \hline
    \multicolumn{1}{|l|}{GITS (Excluding library code)}                     & \multicolumn{1}{l|}{14}                                                                        & 104                                                                      & \multicolumn{1}{l|}{{\color[HTML]{FE0000} \xmark}}                                                  & \multicolumn{1}{l|}{{\color[HTML]{FE0000} \xmark}}                                                 & \multicolumn{1}{l|}{{\color[HTML]{FE0000} \xmark}}                                                        & {\color[HTML]{FE0000} \xmark}                                                  \\ \hline
    \multicolumn{1}{|l|}{CloudSim Express}         & \multicolumn{1}{l|}{4}                                                                          & 11                                                                       & \multicolumn{1}{l|}{{\color[HTML]{009901} \cmark}}                                                   & \multicolumn{1}{l|}{{\color[HTML]{009901} \cmark}}                                                  & \multicolumn{1}{l|}{{\color[HTML]{009901} \cmark}}                                                       & {\color[HTML]{009901} \cmark}                                                 \\ \hline
    \end{tabular}
}
\caption{Productivity Comparison of Scripted Simulation Approaches}
\label{perf-compare-table}
\end{table}

The quantitative values of the approaches are denoted in Table \ref{perf-compare-table}. The CloudSim Plus Automation involves the largest code base in terms of both code complexity and lines of code. This is because the required customization in the use case mandates analyzing the framework code base and identifying how custom classes can be incorporated in the best possible manner, and then recompiling the framework with customization. In comparison, GITS involves a relatively smaller code base. This is because the script-to-java translation is performed via the provided library file, which can be used out of the box. The library file is then consumed in the standard CloudSim simulation implementation, resulting in a relatively smaller code base overall. Among the three approaches, \textit{CloudSim Express} provides the smallest code base as it only requires implementing the custom classes. The extension injecting is done dynamically without involving coding implementations.

The qualitative values of approaches are also denoted in Table \ref{perf-compare-table}. The CloudSim Plus Automation needs to be modified to incorporate the customized classes; thus, it does not support runtime extension injecting, and a framework re-compilation is also needed. However, it does not require CloudSim simulation implementation, as the system model is generated via the provided script with extension support, and the simulation lifetime is managed via the framework itself. On the other hand, GITS performs worst in all qualitative aspects. This is because it only supports converting the script information to Java objects. The rest of the simulation has to be implemented via CloudSim in a standard manner. The \textit{CloudSim Express} provides the best development experience since its dynamic extension injecting feature only requires providing the extension classes. The rest of the simulation is managed automatically.

\begin{figure}[t]
    \centering
    \includegraphics[width=0.5\linewidth]{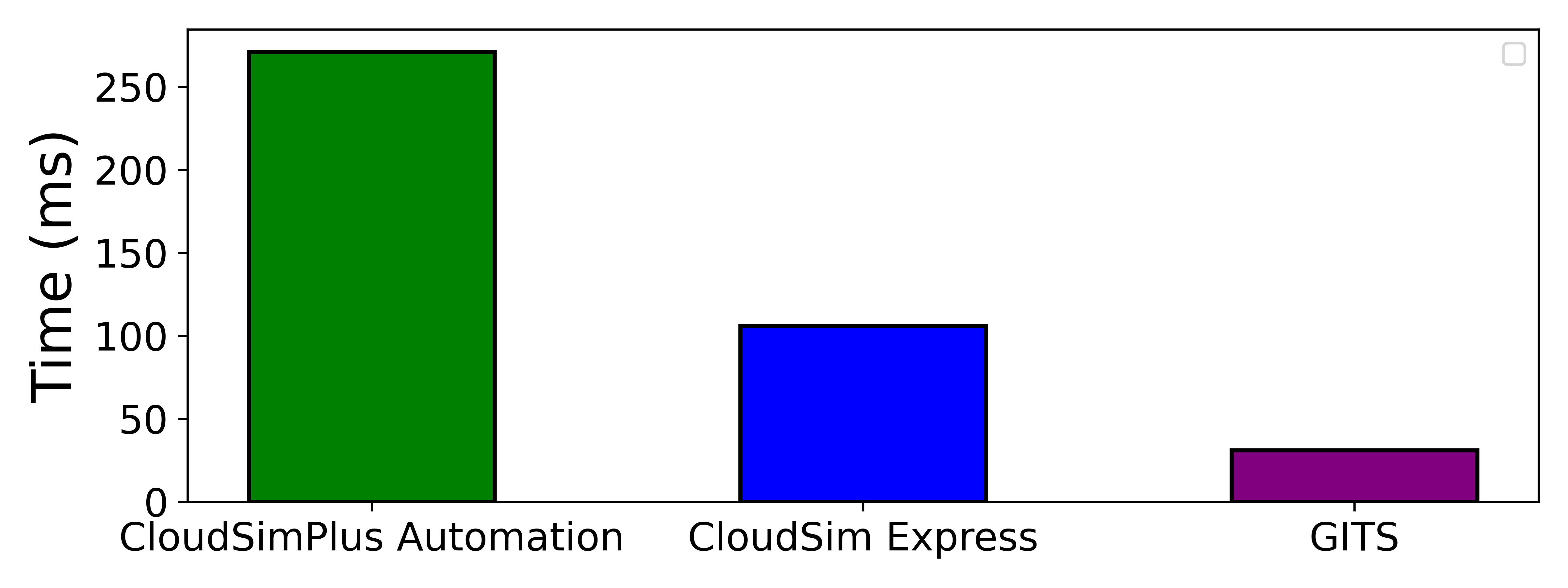}
    \caption{Comparison of Processing Overhead Times}
    \label{fig:interpret-times-comp}
\end{figure}

{%
Fig. \ref{fig:interpret-times-comp} compares the processing overhead times for each approach when handling the \textit{System Model Script}. For the GITS approach, this duration was measured from the tool's launch time to the initiation of the CloudSim simulation, thus encompassing the time taken to process the script and convert it into the model. The method for calculating time is identical for \textit{CloudSim Express}, covering both script processing and framework loading times. However, the latter tends to be lengthier compared to GITS. In contrast, for CloudSimPlus Automation, the measured duration extends from the tool's launch to the invocation of the CloudSimPlus framework, which is the lengthiest of the three. Notably, all approaches incur a processing delay of less than a quarter of a second before commencing the actual simulation, indicating that the impact of the \textit{System Model Script} processing time is generally minimal. Compared to the baseline approaches,\textit{CloudSim Express}  exhibits a reasonable overhead time.
}

To summarize, we observed that \textit{CloudSim Express} outperforms the baselines in terms of productivity, both qualitatively and quantitatively. We evaluated all approaches using a practical simulation use case that emphasized the need for extensibility, which is a prominent feature of the CloudSim toolkit. CloudSim Plus Automation offers a reasonably good developer experience by automating the simulation life cycle through a \textit{System Model Script}  that supports the configuration of different policies at runtime. However, implementing extensions requires managing a relatively large code base and necessitates platform re-compilation. On the other hand, GITS involves a smaller code base but provides only partial automation of the simulation life cycle. \textit{CloudSim Express}  stands out by providing dynamic extension capabilities while automating the simulation life cycle, having a reasonable processing time overhead. In addition to its developer-friendly qualitative aspects, it also achieves a reduction of $71.43$\% in code complexity and $89.42$\% in lines of code compared to the baselines.

\section{Related Work}
\label{section-related-work}

We investigate the literature for approaches that reduce programming language overhead in developing simulated Cloud computing experiments. Cloud simulators are usually bound to a specific programming language \cite{calheiros2011cloudsim, malik2014cloudnetsim++, damian2018score}. Due to that, expertise in programming languages is needed before implementing a simulation scenario, which is an additional overhead. Instead, most approaches offer either textual (in the form of a script) or graphical modeling of Cloud environments \cite{toledo2020epcsac, tsakanikas2022vfcsim, nunez2012icancloud, TeixeiraSá2014cloudreports, li2012dartcsim, wickremasinghe2012cloudanalyst, silva2014cloudsimplusautomation}, or both \cite{jammal2019gits}. Most such approaches allow  implementing end-to-end simulations without coding involved, or with a reduced amount of code \cite{jammal2019gits}. 
Another approach attempts to reduce the effort required for developing scaling simulations by adopting IEEE standards for interoperability \cite{cloudsimscale}. However, it still requires considerable programming effort to implement the scenario. Another proposal suggests simplifying existing simulation approaches by providing flexible combinations of different functional components \cite{easiei}. However, this also requires considerable programming effort.
Productivity can also be improved by the manner that an approach is deployed. There are two primary approaches used in the literature, as a tool and as a library. Most approaches are implemented in the form of a tool \cite{toledo2020epcsac, nunez2012icancloud, TeixeiraSá2014cloudreports, li2012dartcsim, wickremasinghe2012cloudanalyst, silva2014cloudsimplusautomation}, which either a Graphical User Interface (GUI) based \cite{wickremasinghe2012cloudanalyst} or a command line based \cite{silva2014cloudsimplusautomation}. Library-based approaches \cite{jammal2019gits, tsakanikas2022vfcsim} still require partial coding implementation. Alongside that, some approaches are tightly coupled to the underlying simulator. Such implementations mandate the use of a specific simulator implementation \cite{toledo2020epcsac, TeixeiraSá2014cloudreports, jammal2019gits}. In contrast, approaches offered as a separate component from the simulator provide reduced effort in customization \cite{tsakanikas2022vfcsim, nunez2012icancloud, li2012dartcsim, wickremasinghe2012cloudanalyst, silva2014cloudsimplusautomation}. Our proposed approach opts for maximum ease of use, thus implementing a tool that consumes textual representation of the system model. Its generic architecture is implemented as a decoupled layer providing room for customisation.

\par Simplified simulation approaches tend to encapsulate internal simulation details, thus often compromising in customization. For example, in the textual or graphical representation of the system model, only a limited set of Cloud environment elements are available \cite{nunez2012icancloud, TeixeiraSá2014cloudreports}. Introducing a new element, or modifying an existing element involves a complex development effort that often requires framework recompilation.
{
For example, EdgeSimPy, a python based edge computing simulator simplifies scenario definitions via a json input file, however, users are constrained within its schema \cite{edgesimpy}, therefore granular changes require modifications in the simulator code.
Some approaches allow partial customization via mandatory extension points in the simulator platform, such as allocation policies in CloudSim \cite{calheiros2011cloudsim}, providing partial dynamic extension capabilities. For example, DartCSim \cite{li2012dartcsim} allows users to dynamically inject extensions without a framework recompilation by providing an embedded code editor. However, users are limited to the provided set of methods. CloudReports \cite{TeixeiraSá2014cloudreports} further enhance this by allowing users to inject extensions that are developed separately. It still requires providing a separate XML file, which is an overhead. Apart from that, a framework customization might required. For example, the vanilla implementation of CloudSim Plus Automation \cite{silva2014cloudsimplusautomation} does not support allocation policies in a custom Java package, which requires an extension of the framework. In general, framework extensibility is rarely supported (one such case is CloudAnalyst \cite{wickremasinghe2012cloudanalyst}, which supports framework extensibility in its architecture), thus requiring complex code analysis. Our proposed approach supports extensibility in its design by providing a common extension consumption approach for both extension points in the simulator, and the framework itself. Its generic design with handlers for Cloud environment components provides native support for new Cloud environment components, as well as modifying existing ones.

Table \ref{com-related-work} depicts comparison amongst the related work, and the proposed approach. Most approaches significantly compromise between simplified ease of use, and ability to customize. Our proposed approach is designed from its architecture to jointly improve both of those aspects.

\begin{table}[t]
\resizebox{\textwidth}{!}{
\centering
\begin{tblr}{
  cell{2}{2} = {fg=JapaneseLaurel},
  cell{2}{3} = {fg=JapaneseLaurel},
  cell{2}{4} = {fg=JapaneseLaurel},
  cell{2}{5} = {fg=JapaneseLaurel},
  cell{2}{6} = {fg=JapaneseLaurel},
  cell{2}{7} = {fg=red},
  cell{2}{8} = {fg=red},
  cell{3}{2} = {fg=JapaneseLaurel},
  cell{3}{3} = {fg=JapaneseLaurel},
  cell{3}{4} = {fg=WebOrange},
  cell{3}{5} = {fg=WebOrange},
  cell{3}{6} = {fg=red},
  cell{3}{7} = {fg=red},
  cell{3}{8} = {fg=red},
  cell{4}{2} = {fg=JapaneseLaurel},
  cell{4}{3} = {fg=JapaneseLaurel},
  cell{4}{4} = {fg=JapaneseLaurel},
  cell{4}{5} = {fg=WebOrange},
  cell{4}{6} = {fg=red},
  cell{4}{7} = {fg=red},
  cell{4}{8} = {fg=red},
  cell{5}{2} = {fg=JapaneseLaurel},
  cell{5}{3} = {fg=JapaneseLaurel},
  cell{5}{4} = {fg=JapaneseLaurel},
  cell{5}{5} = {fg=JapaneseLaurel},
  cell{5}{6} = {fg=JapaneseLaurel},
  cell{5}{7} = {fg=red},
  cell{5}{8} = {fg=red},
  cell{6}{2} = {fg=JapaneseLaurel},
  cell{6}{3} = {fg=JapaneseLaurel},
  cell{6}{4} = {fg=JapaneseLaurel},
  cell{6}{5} = {fg=WebOrange},
  cell{6}{6} = {fg=JapaneseLaurel},
  cell{6}{7} = {fg=red},
  cell{6}{8} = {fg=red},
  cell{7}{2} = {fg=JapaneseLaurel},
  cell{7}{3} = {fg=JapaneseLaurel},
  cell{7}{4} = {fg=JapaneseLaurel},
  cell{7}{5} = {fg=WebOrange},
  cell{7}{6} = {fg=red},
  cell{7}{7} = {fg=JapaneseLaurel},
  cell{7}{8} = {fg=red},
  cell{8}{2} = {fg=JapaneseLaurel},
  cell{8}{3} = {fg=WebOrange},
  cell{8}{4} = {fg=WebOrange},
  cell{8}{5} = {fg=JapaneseLaurel},
  cell{8}{6} = {fg=red},
  cell{8}{7} = {fg=red},
  cell{8}{8} = {fg=red},
  cell{9}{2} = {fg=JapaneseLaurel},
  cell{9}{3} = {fg=JapaneseLaurel},
  cell{9}{4} = {fg=JapaneseLaurel},
  cell{9}{5} = {fg=WebOrange},
  cell{9}{6} = {fg=red},
  cell{9}{7} = {fg=red},
  cell{9}{8} = {fg=red},
  cell{10}{2} = {fg=JapaneseLaurel},
  cell{10}{3} = {fg=JapaneseLaurel},
  cell{10}{4} = {fg=JapaneseLaurel},
  cell{10}{5} = {fg=JapaneseLaurel},
  cell{10}{6} = {fg=JapaneseLaurel},
  cell{10}{7} = {fg=JapaneseLaurel},
  cell{10}{8} = {fg=JapaneseLaurel},
  hlines,
  vlines,
}
                                                   & {\textbf{Human -readable}\\\textbf{ Cloud Environment}\\\textbf{ Modeling}} & {\textbf{Programming}\\\textbf{ Language }\\\textbf{ Syntax}} & {\textbf{Deployment}\\\textbf{ Mode as a tool}} & {\textbf{De-coupled from}\\\textbf{ Simulation }\\\textbf{ Platform}} & {\textbf{Runtime}\\\textbf{ Extension}\\\textbf{ Injection}} & {\textbf{Native Framework}\\\textbf{Extensibility Support}} & {\textbf{Introducing/}\\\textbf{Modifying}\\\textbf{ Environment}\\\textbf{ Components}} \\
\textbf{EPCSAC \cite{toledo2020epcsac}}                                    & \cmark                                          & \xmark                                                 & \textbf{\cmark}                    & \cmark                                                          & \cmark                                           & \xmark                             & \xmark                                                              \\
\textbf{VFCSIM \cite{tsakanikas2022vfcsim}}                                    & \cmark                                            & \xmark                                                 & \xmark                 & \xmark                                                           & \xmark                                       & \xmark                             & \xmark                                                              \\
\textbf{iCanCloud \cite{nunez2012icancloud}}                                 & \cmark                                          & \xmark                                                 & \textbf{\cmark}                    & \xmark                                                           & \xmark                                       & \xmark                             & \xmark                                                              \\
\textbf{CloudReports \cite{TeixeiraSá2014cloudreports}}                              & \cmark                                          & \xmark                                                 & \textbf{\cmark}                    & \cmark                                                          & \cmark                                           & \xmark                             & \xmark                                                              \\
\textbf{DartCSim \cite{li2012dartcsim}}                                  & \cmark                                          & \xmark                                                 & \textbf{\cmark}                    & \xmark                                                           & \cmark                                           & \xmark                             & \xmark                                                              \\
\textbf{CloudAnalyst \cite{wickremasinghe2012cloudanalyst}}                              & \cmark                                          & \xmark                                                 & \textbf{\cmark}                    & \xmark                                                           & \xmark                                       & \cmark                           & \xmark                                                              \\
\textbf{GITS \cite{jammal2019gits}}                                      & {\cmark}                 & \textbf{Partial}                                              & \xmark                 & \cmark                                                          & \xmark                                       & \xmark                             & \xmark                                                              \\
{\textbf{CloudSim Plus}\\\textbf{ Automation \cite{silva2014cloudsimplusautomation}}}     & \cmark                                            & \xmark                                                 & \textbf{\cmark}                    & \xmark                                                           & \xmark                                       & \xmark                             & \xmark                                                              \\
{\textbf{CloudSim}\textbf{ Express}\\\textbf{ (This Work)}} & \cmark                                            & \xmark                                                 & \textbf{\cmark}                    & \cmark                                                          & \cmark                                           & \cmark                           & \cmark                                                       
\end{tblr}
}
\caption{Comparison of Related Work}
\label{com-related-work}
\end{table}

\section{Conclusions and Future Work}
\label{section-conclusion-future-direction}
Cloud computing environment simulators enable the experimentation of novel infrastructure designs and management approaches with significantly less time, cost, and effort. However, they are tightly coupled to programming language ecosystems, requiring effort in designing, configuring, and programming language expertise. Existing work aims to reduce this overhead but often compromises on simulator extensibility. In this work, we propose an architectural framework for Cloud environment simulators to realize a script-based simulation, aiming to reduce the overhead of programming language expertise while minimizing compromises on extensibility. We implemented the proposed framework for the widely used Cloud simulator, the CloudSim toolkit, and compared it against state-of-the-art baselines for a practical use case. Our evaluations show that \textit{CloudSim Express}  achieves extensible simulations with minimal code compilations and surpasses the baselines with over a $71.43$\% reduction in code complexity and an $89.42$\% reduction in lines of code.

\textbf{Future Work:} The proposed framework segregates simulation logic through a series of handlers. We plan to further reduce the inter-dependency of these handlers to achieve even more isolated implementation logic. Additionally, we intend to develop a graphical user interface using the \textit{System Model Script} .
Furthermore, our proposed framework can be extended to integrate powerful scripting languages other than YAML, such as LUA, to enable even greater flexibility in managing extensions.

\section*{Software Availability}
\label{section:software-info}

The \textit{CloudSim Express} is released as an open-source tool under the GPL-3.0 license. It is available on the following website: \url{https://github.com/Cloudslab/cloudsim-express}, and it will be a part of the upcoming release of CloudSim 6.0.


\begin{thebibliography}{10}

\bibitem{david2022futureofcloud2027}
D.~Smith, ``The future of cloud computing in 2027: From technology to business
  innovation.'' Available at
  \url{https://www.gartner.com/en/doc/768816-the-future-of-cloud-computing-in-2027-from-technology-to-business-innovation}
  (2023/04/26).

\bibitem{mansouri2020cloussimulatorsurvey}
N.~Mansouri, R.~Ghafari, and B.~M.~H. Zade, ``Cloud computing simulators: A
  comprehensive review,'' {\em Simulation Modelling Practice and Theory},
  vol.~104, p.~102144, 2020.

\bibitem{li2012dartcsim}
X.~Li, X.~Jiang, P.~Huang, and K.~Ye, ``Dartcsim: An enhanced user-friendly
  cloud simulation system based on cloudsim with better performance,'' in {\em
  Proceedings of the 2012 IEEE 2nd International Conference on Cloud Computing
  and Intelligence Systems}, pp.~392--396, 2012.

\bibitem{calheiros2011cloudsim}
R.~N. Calheiros, R.~Ranjan, A.~Beloglazov, C.~A.~F. De~Rose, and R.~Buyya,
  ``Cloudsim: a toolkit for modeling and simulation of cloud computing
  environments and evaluation of resource provisioning algorithms,'' {\em
  Software: Practice and Experience}, vol.~41, pp.~23--50, 2011.

\bibitem{malik2014cloudnetsim++}
A.~W. Malik, K.~Bilal, K.~Aziz, D.~Kliazovich, N.~Ghani, S.~U. Khan, and
  R.~Buyya, ``Cloudnetsim++: A toolkit for data center simulations in
  omnet++,'' in {\em Proceedings of the 2014 11th Annual High Capacity Optical
  Networks and Emerging/Enabling Technologies (Photonics for Energy)},
  pp.~104--108, 2014.

\bibitem{damian2018score}
D.~Fernández-Cerero, A.~Fernández-Montes, A.~Jakóbik, J.~Kołodziej, and
  M.~Toro, ``Score: Simulator for cloud optimization of resources and energy
  consumption,'' {\em Simulation Modelling Practice and Theory}, vol.~82,
  pp.~160--173, 2018.

\bibitem{sadowski2018moderncodereviewgoogle}
C.~Sadowski, E.~S\"{o}derberg, L.~Church, M.~Sipko, and A.~Bacchelli, ``Modern
  code review: A case study at google,'' in {\em Proceedings of the 40th
  International Conference on Software Engineering: Software Engineering in
  Practice}, p.~181–190, 2018.

\bibitem{prathanrat2018jupyternotebook}
P.~Prathanrat and C.~Polprasert, ``Performance prediction of jupyter notebook
  in jupyterhub using machine learning,'' in {\em Proceedings of the 2018
  International Conference on Intelligent Informatics and Biomedical Sciences
  (ICIIBMS)}, pp.~157--162, 2018.

\bibitem{tsakanikas2022vfcsim}
V.~Tsakanikas and T.~Dagiuklas, ``Vfcsim: A simulation framework for real-time
  multi-service virtual function chains deployment,'' in {\em Proceedings of
  the GLOBECOM 2022 - 2022 IEEE Global Communications Conference},
  pp.~2644--2649, 2022.

\bibitem{jammal2019gits}
M.~Jammal, H.~Hawilo, A.~Kanso, and A.~Shami, ``Generic input template for
  cloud simulators: A case study of cloudsim,'' {\em Software: Practice and
  Experience}, vol.~49, pp.~720--747, 2019.

\bibitem{silva2014cloudsimplusautomation}
M.~C. Silva~Filho and J.~J. P.~C. Rodrigues, ``Human readable scenario
  specification for automated creation of simulations on cloudsim,'' in {\em
  Proceedings of the Internet of Vehicles -- Technologies and Services},
  pp.~345--356, 2014.

\bibitem{wickremasinghe2012cloudanalyst}
B.~Wickremasinghe, R.~N. Calheiros, and R.~Buyya, ``Cloudanalyst: A
  cloudsim-based visual modeller for analysing cloud computing environments and
  applications,'' in {\em Proceedings of the 2010 24th IEEE International
  Conference on Advanced Information Networking and Applications},
  pp.~446--452, 2010.

\bibitem{nunez2012icancloud}
A.~N{\'u}{\~n}ez, J.~L. V{\'a}zquez-Poletti, A.~C. Caminero, G.~G.
  Casta{\~n}{\'e}, J.~Carretero, and I.~M. Llorente, ``icancloud: A flexible
  and scalable cloud infrastructure simulator,'' {\em Journal of Grid
  Computing}, vol.~10, pp.~185--209, 2012.

\bibitem{toledo2020epcsac}
T.~J.~T. Junior and S.~Bruschi, ``Epcsac - extensible platform for cloud
  scheduling algorithm comparison,'' in {\em Proceedings of the Anais
  Estendidos do XXI Simpósio em Sistemas Computacionais de Alto Desempenho},
  pp.~46--53, 2020.

\bibitem{TeixeiraSá2014cloudreports}
T.~Teixeira~S{\'a}, R.~N. Calheiros, and D.~G. Gomes, {\em CloudReports: An
  Extensible Simulation Tool for Energy-Aware Cloud Computing Environments},
  pp.~127--142.
\newblock Springer International Publishing, 2014.

\bibitem{yaml}
yaml.org, ``Yaml ain't markup language™.'' Available at
  \url{https://yaml.org} (2023/04/30).

\bibitem{mastenbroek2021opendc}
F.~Mastenbroek, G.~Andreadis, S.~Jounaid, W.~Lai, J.~Burley, J.~Bosch, E.~van
  Eyk, L.~Versluis, V.~van Beek, and A.~Iosup, ``Opendc 2.0: Convenient
  modeling and simulation of emerging technologies in cloud datacenters,'' in
  {\em Proceedings of the 2021 IEEE/ACM 21st International Symposium on
  Cluster, Cloud and Internet Computing (CCGrid)}, pp.~455--464, 2021.

\bibitem{kliazovich2012greencloud}
D.~Kliazovich, P.~Bouvry, and S.~U. Khan, ``Greencloud: a packet-level
  simulator of energy-aware cloud computing data centers,'' {\em The Journal of
  Supercomputing}, vol.~62, pp.~1263--1283, 2012.

\bibitem{son2015cloudsimsdn}
J.~Son, A.~V. Dastjerdi, R.~N. Calheiros, X.~Ji, Y.~Yoon, and R.~Buyya,
  ``Cloudsimsdn: Modeling and simulation of software-defined cloud data
  centers,'' in {\em Proceedings of the 2015 15th IEEE/ACM International
  Symposium on Cluster, Cloud and Grid Computing}, pp.~475--484, 2015.

\bibitem{issariyakul2009ns2}
T.~Issariyakul, E.~Hossain, T.~Issariyakul, and E.~Hossain, {\em Introduction
  to network simulator 2 (NS2)}.
\newblock 2009.

\bibitem{cloudsimplus}
M.~C. Silva~Filho, R.~L. Oliveira, C.~C. Monteiro, P.~R.~M. Inácio, and M.~M.
  Freire, ``Cloudsim plus: A cloud computing simulation framework pursuing
  software engineering principles for improved modularity, extensibility and
  correctness,'' in {\em Proceedings of the 2017 IFIP/IEEE Symposium on
  Integrated Network and Service Management (IM)}, pp.~400--406, 2017.

\bibitem{lei2022yamlconfigs}
W.-C. Lei, Y.-P. Chang, and L.-D. Chou, ``Miniwan: A new framework for
  simulating multi-segment network topology based on mininet,'' in {\em
  Proceedings of the 2022 13th International Conference on Information and
  Communication Technology Convergence (ICTC)}, pp.~105--107, 2022.

\bibitem{openapi3}
spec.openapis.org, ``Openapi specification v3.1.0.'' Available at
  \url{https://spec.openapis.org/oas/v3.1.0} (2023/04/30).

\bibitem{romero2023openapi}
J.~Romero-{\'A}lvarez, J.~Alvarado-Valiente, E.~Moguel, J.~Garc{\'i}a-Alonso,
  and J.~M. Murillo, ``Using open api for the development of hybrid
  classical-quantum services,'' in {\em Proceedings of the Service-Oriented
  Computing -- ICSOC 2022 Workshops}, pp.~364--368, 2023.

\bibitem{cloudsimplusautomationgithub}
M.~Campos, ``Cloudsim plus automation: Human-readable scenario specification
  tool for automated creation of simulations on cloudsim and cloudsim plus.''
  Available at \url{https://github.com/cloudsimplus/cloudsimplus-automation}
  (2023/09/19).

\bibitem{silva2017cloudsimplus}
M.~C. Silva~Filho, R.~L. Oliveira, C.~C. Monteiro, P.~R.~M. Inácio, and M.~M.
  Freire, ``Cloudsim plus: A cloud computing simulation framework pursuing
  software engineering principles for improved modularity, extensibility and
  correctness,'' in {\em Proceedings of the 2017 IFIP/IEEE Symposium on
  Integrated Network and Service Management (IM)}, pp.~400--406, 2017.

\bibitem{jaCoCo-counters}
``Jacoco - coverage counter.'' Available at
  \url{https://www.eclemma.org/jacoco/trunk/doc/counters.html} (2023/06/23).

\bibitem{cloudsimscale}
B.~Elahi, A.~W. Malik, A.~U. Rahman, and M.~A. Khan, ``Toward scalable cloud
  data center simulation using high-level architecture,'' {\em Software:
  Practice and Experience}, vol.~50, pp.~827--843, 2020.

\bibitem{easiei}
X.~Su, J.~Qi, J.~Wang, R.~Wang, and Y.~Yao, ``Easiei: A simulator to flexibly
  modeling complex edge computing environments,'' {\em IEEE Internet of Things
  Journal}, vol.~early access, pp.~1--1, 2023.

\bibitem{edgesimpy}
P.~S. Souza, T.~Ferreto, and R.~N. Calheiros, ``Edgesimpy: Python-based
  modeling and simulation of edge computing resource management policies,''
  {\em Future Generation Computer Systems}, vol.~148, pp.~446--459, 2023.

{%
\bibitem{mampage2023cloudsimsc}
A.~Mampage and R.~Buyya, ``Cloudsimsc: A toolkit for modeling and simulation of
  serverless computing environments,'' in {\em Proceedings of the 25th IEEE
  International Conferences on High Performance Computing and Communications (HPCC)},
  2023.

\bibitem{nguyen2023iquantum}
H.~T. Nguyen, M.~Usman, and R.~Buyya, ``iquantum: A case for modeling and
  simulation of quantum computing environments,'' in {\em Proceedings of the
  2nd IEEE International Conference on Quantum Software (QSW)}, 2023.

\bibitem{workflowsim2012chen}
W.~Chen and E.~Deelman, ``Workflowsim: A toolkit for simulating scientific
  workflows in distributed environments,'' in {\em Proceedings of 2012 IEEE 8th
  International Conference on E-Science}, pp.~1--8, 2012.

\bibitem{cloudsimdisk2015louis}
B.~Louis, K.~Mitra, S.~Saguna, and C.~Åhlund, ``Cloudsimdisk: Energy-aware
  storage simulation in cloudsim,'' in {\em Proceedings of 2015 IEEE/ACM 8th
  International Conference on Utility and Cloud Computing (UCC)}, pp.~11--15,
  2015.

\bibitem{networkcloudsim2011garg}
S.~K. Garg and R.~Buyya, ``Networkcloudsim: Modelling parallel applications in
  cloud simulations,'' in {\em Proceedings of 2011 Fourth IEEE International
  Conference on Utility and Cloud Computing}, pp.~105--113, 2011.

\bibitem{cdosim2021fittkau}
F.~Fittkau, S.~Frey, and W.~Hasselbring, ``Cdosim: Simulating cloud deployment
  options for software migration support,'' in {\em Proceedings of 2012 IEEE
  6th International Workshop on the Maintenance and Evolution of
  Service-Oriented and Cloud-Based Systems (MESOCA)}, pp.~37--46, 2012.

\bibitem{cloudsimex}
N.~Grozev, ``Cloudsimex: A set of extensions for the cloudsim simulator.''
  Available at \url{https://github.com/Cloudslab/CloudSimEx} (2023/11/11).

\bibitem{cloud2sim2014kathiravelu}
P.~Kathiravelu and L.~Veiga, ``Concurrent and distributed cloudsim
  simulations,'' in {\em Proceedings of 2014 IEEE 22nd International Symposium
  on Modelling, Analysis \& Simulation of Computer and Telecommunication
    Systems}, pp.~490--493, 2014.


\bibitem{ifogsim22022redowan}
R.~Mahmud, S.~Pallewatta, M.~Goudarzi, and R.~Buyya, ``ifogsim2: An extended
  ifogsim simulator for mobility, clustering, and microservice management in
  edge and fog computing environments,'' {\em Journal of Systems and Software},
  vol.~190, p.~111351, 2022.

}
\end{thebibliography}
\end{document}